\documentclass[sigconf]{acmart}

\usepackage{multirow} 
\usepackage{array}
\usepackage{makecell}
\usepackage{subcaption}
\usepackage{enumitem}

\usepackage{pifont}
\usepackage{threeparttable}

\AtBeginDocument{%
  }
\acmDOI{XXXXXXX.XXXXXXX}

\begin{document}

% \title{Towards Supervised Graph Contrastive Learning on Recommendation: Graph Augmentation via Interpolation Mixup}

\title{Graph Neural Controlled Differential Equations For Collaborative Filtering}

%Unifying the Power of Supervised Siganl and Self-Supervised Training on Graph Recommendation: 


\author{Ke Xu}
\email{kxu25@uic.edu}
\affiliation{%
  \institution{University of Illinois Chicago}
  \city{Chicago}
  \country{USA}}

\author{Weizhi Zhang}
\email{wzhan42@uic.edu}
\affiliation{%
  \institution{University of Illinois Chicago}
  \city{Chicago}
  \country{USA}}

\author{Zihe Song}
\email{zsong29@uic.edu}
\affiliation{%
  \institution{University of Illinois Chicago}
  \city{Chicago}
  \country{USA}}

\author{Yuanjie Zhu}
\email{yzhu224@uic.edu}
\affiliation{%
  \institution{University of Illinois Chicago}
  \city{Chicago}
  \country{USA}}

\author{Philip S. Yu}
\email{psyu@uic.edu}
\affiliation{%
  \institution{University of Illinois Chicago}
  \city{Chicago}
  \country{USA}}


\begin{abstract}

Graph Convolution Networks (GCNs) are widely considered state-of-the-art for recommendation systems. Several studies in the field of recommendation systems have attempted to apply collaborative filtering (CF) into the Neural ODE framework. These studies follow the same idea as LightGCN, which removes the weight matrix or with a discrete weight matrix. However, we argue that weight control is critical for neural ODE-based methods. The importance of weight in creating tailored graph convolution for each node is crucial, and employing a fixed/discrete weight means it cannot adjust over time within the ODE function. This rigidity in the graph convolution reduces its adaptability, consequently hindering the performance of recommendations. In this study, to create an optimal control for Neural ODE-based recommendation, we introduce a new method called Graph Neural Controlled Differential Equations for Collaborative Filtering (CDE-CF). Our method improves the performance of the Graph ODE-based method by incorporating weight control in a continuous manner. To evaluate our approach, we conducted experiments on various datasets. The results show that our method surpasses competing baselines, including GCNs-based models and state-of-the-art Graph ODE-based methods.
\end{abstract}


% \ccsdesc[500]{Information systems~Recommender systems}
% \ccsdesc[500]{Information systems~Collaborative filtering}
% \ccsdesc[300]{Mathematics of computing~Ordinary differential equations}





\maketitle

\section{Introduction}
In recent years, there has been a surge in the popularity of Graph Convolutional Networks (GCNs) \cite{zhou_graph_2020} for machine learning tasks involving graph data. GCNs have gained popularity in recent years for their effectiveness in learning node embeddings by exploiting the structure of the graph. 
Collaborative Filtering (CF) \cite{he_neural_2017, zhang2023dual} is a popular approach in recommender systems. Since GCNs can effectively capture relationships between users and items in a graph, they have shown promising results in improving the performance of collaborative filtering, particularly in scenarios with complex and sparse user-item interaction data.

Several studies \cite{he_lightgcn:_2020, zhang2024mixed, zhang2024we} have shown that linear GCN architectures outperform non-linear ones for collaborative filtering \cite{he_neural_2017}. Moreover, linear GCNs can be easily interpreted as an ordinary differential equation. This concept \cite{Chen_node, deng_continuous_2019} has led to the development of LT-OCF, a Neural Ordinary Differential Equations (NODEs)-based CF method \cite{choi_lt-ocf:_2021}. LT-OCF demonstrates the suitability of NODEs-based approaches for collaborative filtering. The main idea behind LT-OCF is to create a continuous version of the GCN layer, resembling LightGCN \cite{he_lightgcn:_2020} but with a customizable number of layers.

GODE-CF \cite{xu_graph_2023} is another NODE-based method for collaborative filtering. Inspired by of Graph Neural Ordinary Differential Equations (GODEs)\cite{poli_graph_2021}, instead of creating a continuous message-passing layer, GODE-CF parametrizes the ODE using one or two GCN layers. It tries to utilize the information captured by these GCN layers to estimate the final state of the embedding by solving an ODE problem. Unlike LT-OCF \cite{choi_lt-ocf:_2021}, GODE-CF incorporates a discrete weight for each node embedding. However, it remains unclear whether the weight is helpful or not. The experimental results in GODE-CF indicate that the weight does not always improve performance across all cases. We argue that a discrete weight may also limit the performance of GODE-CF. ODE is a continuous form, and incorporating a weight matrix in a continuous manner should further enhance the performance of GODE-CF. 

Motivated by this idea and based on the framework of GODE-CF, we propose a new method called Graph Neural Controlled Differential Equations for Collaborative Filtering (CDE-CF). Our method is based on the framework of GODE-CF. However, unlike GODE-CF, we incorporate MLPs to control the ODE instead of using a discrete weight for each node. Such MLP weight generator can be regarded as part of the ODE function producing continuous weight values for the continuous time slots. To evaluate the performance of our method, we use four public review datasets and compare them with state-of-the-art methods.
The results indicate that our method consistently outperforms the mentioned methods in all datasets. Moreover, we showcase the efficiency of our method by demonstrating its faster training compared to most GCNs-based methods. Furthermore, we explore the influence of different ODE solvers on various datasets. To encourage future exploration of CDE-CF, we have made our work open-source on \textcolor{blue}{\url{https://github.com/DavidZWZ/CDE-CF}}. In summary, our contributions can be outlined as follows:
\begin{itemize}[leftmargin=*]
    \item We identify the limitations of non-weight and discrete weight in graph-based collaborative filtering and demonstrate the importance of the control matrix in ODE-based methods.
    \item We have developed ODE-CF, a novel method that can adaptively control the weight along the time and for different nodes in the Graph Neural ODE function for collaborative filtering.
    \item We conduct extensive experiments on four real-world datasets to test the effectiveness of CDE-CF. It achieves the highest performance with the same training time as CDE-CF, demonstrating the remarkable performance of CDE-CF.
\end{itemize}

\section{Preliminaries}
\textbf{Neural Ordinary Differential Equations (NODEs)} refers to a method that is used to model the continuous dynamics of hidden states within neural networks \cite{Chen_node}. This is achieved by characterizing the dynamics through an ordinary differential equation (ODE) that is parameterized by a neural network. The main objective of this method is to learn implicit differential equations from data. By employing neural networks to parameterize the ODEs, it becomes possible to capture intricate patterns in the data that would be difficult to capture using discrete methods. NODEs offer a framework for modeling complex systems by leveraging ODEs to capture continuous behavior. The formula for NODEs is written as follows:
\begin{equation}
    h(t_1) = h(t_0) + \int_{t_0}^{t_1} f(h(t), \theta )dt
\end{equation}
where $f$ is a neural network parameterized by $\theta$ that approximates $\frac{dh(t)}{dt}$. This approximation allows us to derive $h(t_1)$ from $h(t_0)$. The parameter $\theta$ is trained using data. The variables $t_0$ and $t_1$ represent the starting and ending times, with $t_0$ often set to 0. In Neural ODEs, $t_1$ can be considered as the number of layers in a neural network.

% \subsection{Graph Neural Ordinary Differential Equations (GODEs)}
% The concept behind Graph Neural Ordinary Differential Equations (GODEs) \cite{poli_graph_2021} is to parametrize the derivative function using 2 or 3 GCN layers. This concept incorporates ideas from graph spectral theory and Neural ODEs. Based on the graph spectral theory, the residual version of GCN layers \cite{kipf_semi-supervised_2017,Kenta_graph} takes the following form:
% \begin{equation}
%     E_{n+1} = \sigma(AGG(W, A, E_n))
% \end{equation}
% where $A$ is the adjacency matrix and $\sigma$ is a nonlinear activation function, $W$ is the transformation matrix, and $E_{n+1}$ is the node embedding at the $(n+1)$th layer. And the GODEs to compute the final node embeddings is:
% \begin{equation}
%     E(t_1) = E(0) + \int^{t_1}_{t_0} GCN(n, E(t)) dt
% \end{equation}
% where $E(t)$ represents the node embeddings at time $t$, $E(0)$ represents the initial embeddings, and $GCN(n,E(t))$ represents the $n$ graph convolution layer with input node embeddings $E(t)$. The final node embeddings, denoted as $E(t_1)$, are estimated by the ODE using $E(t)$ as input. When working with GODEs, higher-order ODE solvers are generally more effective, particularly when the graph is dense enough to benefit from the additional computations \cite{poli_graph_2021}.

\begin{figure}[htpb]
\centering
\includegraphics[width = 1.0\linewidth]{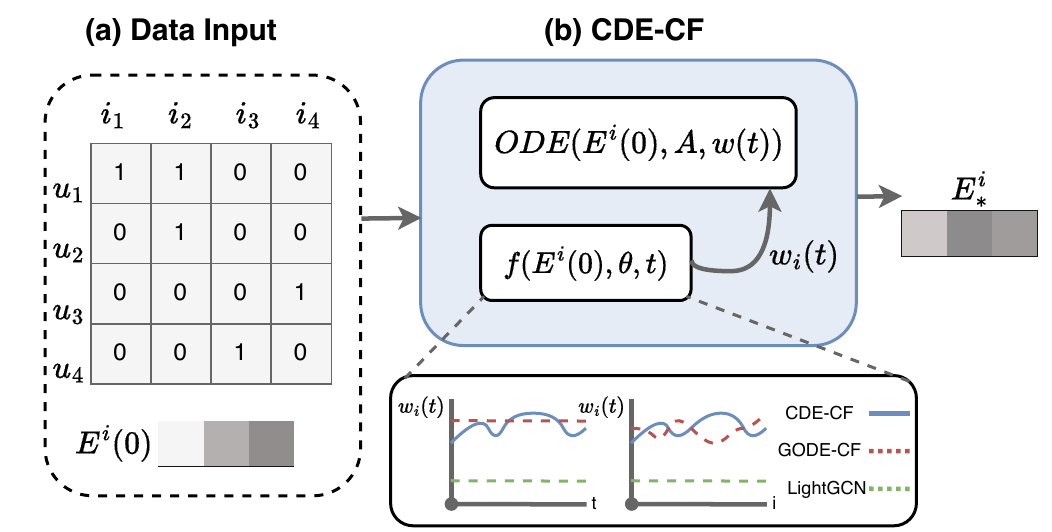}
\caption{The graph convolution process for CDE-CF to generate an item final embedding $E^i_*$ and the user embedding $E^u_*$ can be produced in a similar process. (a) The data input contains the bipartite adjacency matrix and initial embedding. (b) The architecture of the proposed CDE-CF includes the ODE function to model the graph convolution and the weight generator to produce continuous weight. The plot below compares the weight value of CDE-CF, GODE-CF, and LightGCN with the change of time $t$ and the node index $i$.}
\label{fig: main}
\end{figure}

\section{Proposed Method}
Graph Neural Ordinary Differential Equations-based method for Collaborative Filtering (GODE-CF) \cite{xu_graph_2023} is a method that draws inspiration from the concept of Graph-based NODEs \cite{poli_graph_2021}. Instead of creating a continuous message-passing layer, GODE-CF directly parameterizes the derivative function using one or two layers of GCNs. In other words, GODE-CF utilizes the information captured by two LightGCN layers to estimate the final state of the embedding by solving an ODE problem.

\begin{table*}[htpb]
\small
    \centering
    \caption{Overall performance of CDE-CF in comparison with different state-of-the-art baselines on four datasets.}
    \begin{tabular}{c|cc|cc|cc|cc}
    \hline
        Dataset &  \multicolumn{2}{c}{Beauty} &  \multicolumn{2}{c}{Health} &  \multicolumn{2}{c}{Cell Phone} &  \multicolumn{2}{c}{Office} \\ \hline
        Method & Recall@20 & NDCG@20 & Recall@20 & NDCG@20 & Recall@20 & NDCG@20 & Recall@20 & NDCG@20 \\ \hline        
        NGCF & 0.07079 & 0.02995 & 0.03064 & 0.01226 & 0.04387 & 0.01691 & 0.05097 & 0.022137 \\ 
        layerGCN & 0.07620 & 0.03144 & 0.02453 & 0.01009 & 0.03967 & 0.01508 & 0.04710 & 0.02105 \\ 
        UltraGCN & 0.05661 & 0.02618 & 0.03170 & 0.01348 & 0.03604 & 0.01558 & 0.04689 & 0.02336 \\ 
        GTN & 0.07146 & 0.03059 & 0.03287 & 0.01351 & 0.04559 & 0.01780 & 0.04363 & 0.02026 \\ 
        LightGCN & 0.07776 & 0.03299 & 0.03030 & 0.01202 & 0.04429 & 0.01660 & 0.05647 & 0.02635 \\ 
        LT-OCF & 0.07879 & 0.03309 & 0.03022 & 0.01197 & 0.04641 & 0.01739 & 0.05626 & 0.02596 \\    
        GODE-CF & 0.08075 & 0.03406 & 0.03387 & 0.01356 & 0.05079 & 0.01909 & 0.05667 & 0.02702 \\ 
        \hline
        CDE-CF & \textbf{0.08129} & \textbf{0.03426} & \textbf{0.03468} & \textbf{0.01375} & \textbf{0.05082} & \textbf{0.01945} & \textbf{0.05728} & \textbf{0.02713}\\
        \hline
    \end{tabular}
    \label{Table:overall}
\end{table*}

Differnt from LightGCN \cite{he_lightgcn:_2020}, GODE-CF does not involve layer combinations, as the integration can be viewed as the summation of all layers from time 0 to $t_1$. The initial embeddings serve as the input for the ODE, and the output of the ODE becomes the final embedding. The overall formula can be expressed as:
\begin{equation}
\begin{split}
    E^u_{*} &= E^u(0) + \int^{t_1}_{0} W(A^n-I)E^i(t) dt\\
    E^i_{*} &= E^i(0) + \int^{t_1}_{0} W(A^n-I)E^u(t) dt
\end{split}
\end{equation}
where the normalized adjacency matrix is denoted by $A$, and the initial user and item embeddings are represented by $E^u_0$ and $E^i_0$, respectively. The discrete weight matrix is denoted by $W$, and $n$ represents the number of layers. The final user embeddings and item embeddings are denoted as $E^u_{*}$ and $E^i_{*}$, respectively.

Unlike typical GCN-based models \cite{he_neural_2017,he_lightgcn:_2020}, which combine the embeddings from all layers, GODE-CF estimates the final embeddings by leveraging information from multiple GCN layers through an ODE function. Similar to other methods, the embeddings will be trained using the BPR \cite{rendle_bpr:_2009} loss and ODE solvers like explicit Euler and RK4 will be used to solve the ODE problem.

Distinct from LightGCN, which removes the weight matrix, ODE-based methods rely on a weight matrix to regulate the progression of each node toward its optimal state. Without weights, all node embeddings would converge in the same state at the the same timestep, resulting in suboptimal embeddings for some nodes. Since nodes may require different timesteps to reach their optimal states, weights play a crucial role in ODE-based methods. GODE-CF introduces a discrete weight that improves performance in specific scenarios. However, this method has limitations due to the continuous nature of the overall framework. We argue that incorporating weights in a continuous manner would further enhance the performance of GODE-CF.

Instead of simply creating a weight matrix like GODE-CF, we propose GODE-CF and build MLPs to control the ODE. This ensures that each node reaches its optimal state. The idea is quite simple. Fig ~\ref{fig: main} demonstrates ODE-based modeling for graph convolution and a controller to produce the weight varying from the time $t$ and the node index $i$. We will use the initial embedding as the input for the MLPs, and the output of the MLPs will serve as the weight matrix. And, we integrate the MLPs into the ODE framework. The whole framework can be treated as two parts: (1) \textit{An Neural ODE with the initial node embeddings as the input to estimate the weight matrix.} (2) \textit{A Graph Neural ODE with initial node embeddings and weight matrix to estimate the final embedding.} The first component is to control the ODE that ensures each node will reach an optimal embedding. We combine these two parts into one single ODE function. The overall formula can be written as follows:
\begin{equation}
\begin{split}
    E^u_{*} &= E^u(0) + \int^{t_1}_{0} \sigma(f(E^u(t), \theta)) (A^n-I)E^i(t) dt\\
    E^i_{*} &= E^i(0) + \int^{t_1}_{0} \sigma(f(E^i(t), \theta)) (A^n-I)E^u(t) dt
\end{split}
\end{equation}
where $E_{*}^{u}$ is the final users embeddings and $E_{*}^i$ is the final items embeddings. $f(E^u(t), \theta)$ and $f(E^i(t), \theta)$ represent the MLPs with the user embedding and item embedding at time step $t$, respectively. Here, $\theta$ represents the parameters of the MLPs. Additionally, $\sigma$ denotes the sigmoid function. The output of the MLPs is the weight matrix at time step $t$. The final user embeddings and item embeddings are denoted as $E^u_{*}$ and $E^i_{*}$, respectively. Similar to GODE-CF, we employ the BPR loss for training the embeddings and ODE solvers, such as Euler or RK4, to solve the ODE. To make predictions, we follow the same settings as GODE-CF. Once we obtain the final embeddings, the prediction is calculated as the inner product of the user embeddings and item embeddings:
$
    y_{u,i} = E_{*}^{u^T} E_{*}^i.
$

\begin{table}[htpb]
\small
    \centering
    \caption{The statistics of the datasets}
    \begin{tabular}{c|c|c|c|c}
    \hline
        Datasets & Training & Validation & Testing & Sparsity \\ \hline
        Office & 43,448 & 4,905 & 4,905 & 0.44867\% \\ 
        Health & 269,137 & 38,609 & 38,609 & 0.0484\% \\ 
        Cell Phone & 138,681 & 27,879 & 27,879 & 0.0668\% \\ 
        Beauty & 153,776 & 22,363 & 22,363 & 0.07335\% \\ \hline
    \end{tabular}
    \label{table:data}
\end{table}

\section{Experiment}

\subsubsection{\textbf{Datasets.}}
We use the public Amazon Reviews dataset \cite{mcauley_image-based_2015} with four benchmark categories, including: \textit{Beauty}, \textit{Health}, \textit{Cell Phones}, \textit{Office Product}. The details of the datasets are summarized in Table~\ref{table:data}. We follow the 5-core setting as existing works on users and the same transformation \cite{he_neural_2017,he_lightgcn:_2020,fan_graph_2022, fan_graph_2023} of treating the existence of reviews as positives. We sort each user’s interactions chronologically and adopt the leave-one-out setting, with the last interacted item for testing and the second last interaction for validation.

\subsubsection{\textbf{Baselines}}
In total, we compare CDE-CF with various types of the state-of-the-art models:
\begin{itemize}[leftmargin=10pt]
\item layerGCN \cite{zhou_layer-refined_2022} is a GCN-based CF method with layer-refinement.
\item LightGCN \cite{he_lightgcn:_2020} is a lightweight linear GCN-based CF method.
\item UltraGCN \cite{mao_ultragcn:_2021} is an ultra-simplified formulation of GCN that directly approximates the limit of infinite message-passing layers.
\item GTN \cite{fan_graph_2022} is a graph trend filtering network framework to capture the adaptive reliability of the interactions. 
\item LT-OCF \cite{choi_lt-ocf:_2021} is a NODE-based method that aims to learn the optimal architecture of the model for graph-based CF.
\item GODE-CF \cite{xu_graph_2023} is a GODE-based method that uses two GCN layers of information to estimate the final embeddings.
\end{itemize}

\subsubsection{\textbf{Evaluation Metrics}}
For the evaluation metrics, Recall@K and NDCG@K are adopted for a fair comparison of all the baselines in the top-K recommendation task. K is set as 20 in the main performance evaluation and is set to 20 by default in the other experiments. The full-ranking strategy is adopted for all the experimental studies, i.e., all the candidate items not interacted with the user will be ranked in testing.

\subsection{Overall Performance Comparison}

\begin{table*}[ht]
\small
    \centering
    \caption{The impact on weight, discrete weight on Beauty, Health, Cell Phone, and Office datasets.}
    \begin{tabular}{lll|ll|ll|ll}
    \hline
        Dataset &  \multicolumn{2}{c}{Beauty} &  \multicolumn{2}{c}{Health} &  \multicolumn{2}{c}{Cell Phone} &  \multicolumn{2}{c}{Office} \\ \hline
        ~ & Recall@20 & NDCG@20 & Recall@20 & NDCG@20 & Recall@20 & NDCG@20 & Recall@20 & NDCG@20 \\ \hline
        Without weight (W) & 0.08076 & 0.03407 & 0.03403 & 0.01353 & 0.05058 & 0.01916 & 0.05586 & 0.02661 \\ 
        With discrete weight (W) & 0.08027 & 0.03395 & 0.03362 & 0.01356 & 0.05079 & 0.01909 & 0.05668 & 0.02702 \\ \hline
        CDE-CF & \textbf{0.08129} & \textbf{0.03426} & \textbf{0.03468} & \textbf{0.01375} & \textbf{0.05082} & \textbf{0.01945} & \textbf{0.05728} & \textbf{0.02713}\\ \hline
    \end{tabular}
    \label{table:weight}
\end{table*}

In this comprehensive experimental study, we evaluated the performance of several state-of-the-art GCN-based methods and ODE-based methods on four diverse datasets. We use Recall@20 and NDCG@20 to measure their performance. Table (\ref{Table:overall}) is the overall performance, and we summarize the main results:

\begin{itemize}[leftmargin=10pt]
\item Predominantly, CDE-CF achieved the highest NDCG@20 and Recall@20 scores across all datasets, highlighting its superior efficacy in recommendation tasks. In the ODE-based method, GODE-CF significantly outperforms the strongest GCN-based baselines. For other baselines, LightGCN exhibits the best performance on \textit{Beauty}, while GTN demonstrates the best performance on \textit{Health} compared to other GCN-based baselines.  
\item Among all baselines, the ODE-based model GODE-CF shows state-of-the-art performance for all cases. This indicates the superiority of the ODE-based methods for modeling the high-order relationship in the graph convolution.
\end{itemize}

\begin{table}[ht]
\small
    \centering
    \caption{Efficiency comparison with LightGCN, LT-OCF, and GODE-CF on four datasets, with 1000 total training epochs.}
    \begin{tabular}{ccccc}
    \hline
        ~ & ~ & Training Time & ~ & ~ \\ \hline
        Dataset & Beauty & Health & Cell Phone & Office \\ \hline
        
        LightGCN & 2393.25s & 4892.29s & 2192.09s & \textbf{382.89s} \\ 
        LT-OCF & 5876.53s & 15785.80s & 5614.14s & 771.50s \\ 
        GODE-CF & 2120.92s & 4823.32s & 2136.18s & 406.28s \\ \hline
        CDE-CF & \textbf{2109.52s} & \textbf{4767.27s} & \textbf{2106.83s} & 421.46s  \\\hline
    \end{tabular}
    \label{table:efficiency}
\end{table}

\section{Ablation Study}
\textbf{Impact on weight.}
Here, we present the comprehensive ablation study of weight components of the CDE-CF model. From the detailed analysis provided in Table (\ref{table:weight}), we observe interesting insights. Firstly, when incorporating a discrete weight, GODE-CF does not consistently outperform GODE-CF without the weight, indicating that a discrete weight alone may not be sufficient to effectively control the ODE system. In comparison, our method CDE-CF, which incorporates continuous weights, surpasses GODE-CF without weight in all cases. This indicates the importance of incorporating continuous weight control in the ODE methods.\vspace{3pt}
\\ 
\textbf{Efficiency Comparison.}
We provide empirical evidence demonstrating the superiority of CDE-CF in terms of training efficiency compared to other baselines. We train all models with a fixed number of 1000 epochs to eliminate the effect of varying epoch numbers. 
As in Table~\ref{table:efficiency}, though CDE-CF includes an additional component compared to GODE-CF, and CDE-CF has a faster training time than GODE-CF in three datasets. The time step $t$ is the main factor that affects the training speed. We have found that the optimal value for $t$ across all cases is approximately $8.5$ for GODE-CF. For our method, the optimal $t$ value is around $6.5$, resulting in a reduced amount of time required to solve the ODE. 

\section{Conclusion}
In this study, we propose a new method called CDE-CF based on the GODE-CF. In particular, a novel control weight is devised to cater to the continuous time in the ODE functions in order to better model the graph convolution process. The experimental results on four different real-world demonstrate that CDE-CF outperforms various state-of-the-art baselines in terms of performance, while also having a shorter training time compared to GODE-CF. Additionally, the ablation study reveals that we have created a more reasonable weight matrix control compared to GODE-CF. For further study, we will explore more complex controllers for ODE-based methods. In conclusion, having a good controller is crucial to further enhance the performance of the ODE-based method and contribute to performance improvements.

\bibliographystyle{ACM-Reference-Format}
\bibliography{sample-base}

\end{document}